\documentclass{iau} 
\usepackage{graphicx}

\title[SWAG Water Masers] %% give here short title %%
{SWAG Water Masers in the Galactic Center}

\author[J\"urgen Ott et al.]   %% give here short author list %%
{J\"urgen Ott$^1$, 
Nico Krieger$^2$, 
Matthew Rickert$^{3}$, 
David Meier$^{4}$, 
Adam Ginsburg$^{5}$,
Farhad Yusef-Zadeh$^{6}$,
 \and the SWAG team}

\affiliation{
$^1$ National Radio Astronomy Observatory, \\ 1003 Lopezville Road,
Socorro, NM 87801, USA \\ email: {\tt jott@nrao.edu} \\[\affilskip]
$^2$ Max-Planck-Institut f\"ur Astronomie, \\ K\"onigstuhl 17, 69120
Heidelberg, Germany \\ email: {\tt krieger@mpia.de}  \\[\affilskip]
$^3$ Department of Physics and Astronomy and CIERA, Northwestern
University, \\ Evanston, IL 60208, USA; Matthew Rickert is a Reber Fellow at the National Radio Astronomy Observatory. \\[\affilskip]
$^4$ New Mexico Institute of Mining and Technology, \\ 801 Leroy Place,
Socorro, NM 87801, USA \\ email {\tt
  David.Meier@nmt.edu}; David S. Meier is also an Adjunct Astronomer at the National Radio Astronomy Observatory.\\[\affilskip]
$^{5}$ National Radio Astronomy Observatory, \\ 1003 Lopezville Road,
Socorro, NM 87801, USA \\ email: {\tt aginsbur@nrao.edu}\\[\affilskip]
$^6$ Department of Physics and Astronomy and CIERA, Northwestern
University, \\ Evanston, IL 60208, USA \\ email: {\tt zadeh@northwestern.edu}
}

\pubyear{2017}
\volume{336}  %% insert here IAU Symposium No.
\jname{Astrophysical Masers: Unlocking the Mysteries of the Universe}
\editors{A. Tarchi, M.J. Reid \& P. Castangia, eds.}
\begin{document}

\maketitle

\begin{abstract}
  The Galactic Center contains large amounts of molecular and ionized
  gas as well as a plethora of energetic objects. Water masers are an
  extinction-insensitive probe for star formation and thus ideal for
  studies of star formation stages in this highly obscured
  region. With the Australia Telescope Compact Array, we observed
  22\,GHz water masers in the
  entire Central Molecular Zone with sub-parsec resolution as part of the
  large SWAG survey: ``Survey of Water and Ammonia in the Galactic
  Center''. We detect of order 600 22\,GHz
  masers with isotropic luminosities down to
  $\sim 10^{-7}$\,L$_{\odot}$. Masers with luminosities of
  $\gtrsim10^{-6}$\,L$_{\odot}$ are likely associated with young
  stellar objects. They appear to be close to molecular gas
  streamers and may be due to star formation events that are triggered
  at pericenter passages near Sgr\,A*. Weaker masers are more widely
  distributed and frequently show double line features, a tell-tale
  sign for an origin in evolved star envelopes.
  
\end{abstract}

\vspace{-0.2cm}
\firstsection % if your document starts with a section,
\section{Introduction}

The Central Molecular Zone (CMZ; the inner $\sim500$\,pc) of the Milky Way is the nearest nucleus of any galaxy and
allows us to study aspects of star formation processes under extreme
conditions, such as high pressure gas, extreme tidal forces, strong
radiation fields, and cloud-cloud shock zones. Gas flows from the
Milky Way disk to the CMZ, where some of the gas appears to follow
specific trajectories also known as orbits or streamers. Kinematic
models of the streamers are given, e.g., in \cite{mol11},
\cite{kru15}, or \cite{rid17}. \cite{kru15}, in particular, predict
that streamers near the pericenter to Sgr\,A* (which marks the center
of the gravitational potential) may compress the gas and initiate a
collapse and trigger star formation. \cite{lon13} and \cite{hen16}
show that indeed a star formation sequence can be observed in the CMZ,
where a streamer after its Sgr\,A* pericenter passage contains dense
gas with no obvious star formation (the ``brick''), followed by
consecutively more evolved star forming regions, culminating in
Sgr\,B2, the most vigorous star formation site in the Milky
Way. Downstream from Sgr B2, stellar clusters appear even more
evolved, e.g. in the radio continuum-bright Sgr B1 region and further down
in the Arches and Quintuplet stellar clusters. \cite{kri17} and
\cite{gin16} also find
evidence that gas near pericenter passages exhibit positive
temperature gradients, which
suggests that gas may indeed be compressed near those spots.

The CMZ is characterized by extreme optical extinction and radio lines
are ideal to derive the evolutionary status of star forming regions
along the possible star formation sequence. 22\,GHz water
($6_{16}-5_{23}$) masers are not affected by extinction and are
produced in extreme environments, typically in shocked envelopes of
evolved stars, outflows of young stellar objects (YSOs), or, in their
most extreme form, in the accretion disks and jet-gas interfaces of
active galactic nuclei (AGN). Maser luminosities from the different
sources vary, where AGN related masers are extremely bright megamasers
(e.g., \cite[Lo 2005]{lo05}; where the term 'megamasers' refers to a
comparison with typical maser strengths of individual stellar sources
in the Milky Way). Masers near evolved stars and YSOs are much less
luminous and the most luminous source in the Milky Way reaches
isotropic luminosities of $\sim 0.1$\,L$_\odot$ (W 49N,
\cite[Liljestrom \etal\ 1989]{lil89}). \cite{pal93} compare maser
luminosities in the Milky Way and they find that water masers related
to YSOs are on the high end of the luminosity function, whereas
evolved stars only reach maximum luminosities of
$\sim 10^{-4}$\,L$_\odot$. With luminosity cutoffs it is therefore
possible to preferentially select YSOs and thus locate the related
active zones of star formation.

\vspace{-0.2cm}
\section{SWAG: Survey of Water and Ammonia in the Galactic Center}
The SWAG survey ``Survey of Water and Ammonia in the Galactic Center''
is ideally suited to obtain a rather comprehensive picture of the
molecular gas toward the Galactic Center. This three year survey with
the Australia Telescope Compact Array\footnote{The Australia Telescope
  Compact Array is part of the Australia Telescope National Facility
  (ATNF), a division of the Commonwealth Scientific and Industrial
  Research Organisation (CSIRO).}, covers the entire CMZ in the
21.2-25.4\,GHz frequency range with high spectral resolution of
targeted 42 specific lines. This includes the 22\,GHz water maser
line, multiple transitions of the temperature tracer ammonia,
photon-dominated and shock-dominated region tracers, and radio
recombination lines. The resolution of SWAG is about $\sim27''$ which
corresponds to sub-pc resolution at the distance of the CMZ
(8.5\,kpc). First results of SWAG are described in \cite{kri17}, who
analyze the temperature properties along the gas streamers based on
multiple transitions of ammonia. Here we report on first results of
the 22\,GHz water maser ($6_{16}-5_{23}$) transition.

\begin{figure}[h]
% \vspace*{-2.0 cm}
\begin{center}
 \includegraphics[width=13cm]{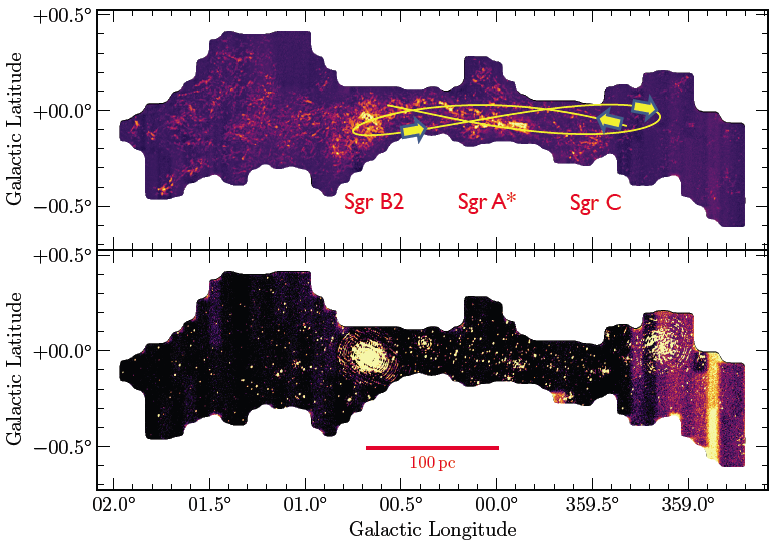} 
% \vspace*{-1.0 cm}
 \caption{{\bf Top:} Peak flux map of the ammonia (3,3) emission of
   the SWAG survey. Prominent features and the gas streamers are
   marked. The arrows indicate the direction of the flows. {\bf
     Bottom:} The peak flux map of the 22\,GHz water masers in the same
   region. Note that some extremely bright masers, in particular in
   the Sgr\,B2 region show considerable sidelobes. }
   \label{fig1}
\end{center}
\end{figure}

\begin{figure}[h]
% \vspace*{-2.0 cm}
\begin{center}
 \includegraphics[width=13cm]{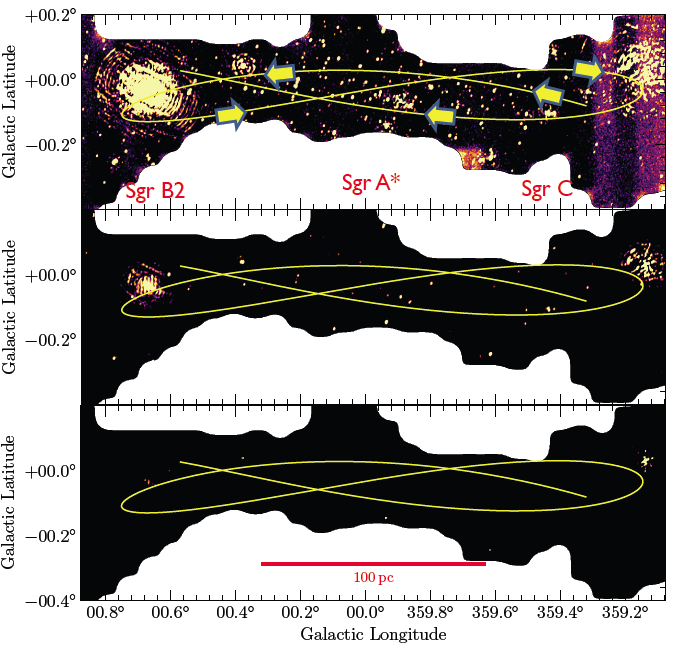} 
% \vspace*{-1.0 cm}
 \caption{Water maser distribution at different isotropic luminosity
   cuts, assuming that all masers are at a distance of 8.5\,kpc. {\bf
     Top:} $>10^{-7}$\,L$_\odot$ {\bf Middle:} $>10^{-6}$\,L$_\odot$
   {\bf Bottom:} $>10^{-5}$\,L$_\odot$. Overlaid are the Streamers and
   directional arrows. }
   \label{fig2}
\end{center}
\end{figure}

\vspace{-0.2cm}
\section{Populations of Water Masers toward the CMZ}
Maps of ammonia and water masers are shown in Fig.\,\ref{fig1}. The
gas streamers mostly reside in the inner $\sim 150$\,pc, the region
that covers the area from Sgr\,B2 to Sgr\,C, roughly centered on the
supermassive black hole Sgr\,A*. In Fig.\,\ref{fig1}, we plot the
location and direction of the \cite{kru15} streamers on top of the gas
distribution. Starting between Sgr\,C and Sgr\,A*, the streamer passes
the first pericenter near Sgr\,A* and continues toward and beyond Sgr
B2 (for a 3-dimensional picture, see \cite[Kruijssen, Dale, \&
Longmore 2015]{kru15}).

The bottom of Fig.\,\ref{fig1} shows the peak fluxes of the 22\,GHz
water masers. Note that some masers, especially near Sgr B2 and
around $b\sim-0.8^{\circ}$, are spilling beam sidelobes
across many pointings. We consider those regions saturated and of
limited use for the current analysis. To first order, the gas and the 22\,GHz
masers are not particularly correlated and a close inspection of the
data shows that even on smaller scales masers and molecular clumps are
not necessarily co-spatial. 

In Fig.\,\ref{fig2}, we show masers in the inner region of the CMZ, spanning
Sgr\,B2 to Sgr\,C, at different
luminosity cuts (assuming that all water masers are at 8.5\,kpc
distance). The most luminous sources with isotropic luminosities
exceeding $10^{-6}$\,L$_\odot$ are indeed close to the streamer
trajectories. Following the work of \cite{pal93}, their luminosities
are consistent with being related to YSOs and therefore trace current
star formation sites. The gas and maser velocities are also frequently
separated by less than $\sim \pm 20$\,km\,s$^{-1}$, which further
supports this scenario. Following the streamers, there is only one
weak water maser source in the ``brick'' at
$(l,b)\sim (0.253^{\circ},0.016^{\circ})$ followed by a very bright
$>10^{-5}$\,L$_\odot$ source near $(l,b)\sim (0.38^{\circ},
0.04^{\circ})$; source ``C'', see
\cite[Ginsburg \etal\ 2015]{gin15}) and the extreme Sgr\,B2 water
masers. A few fainter masers but with luminosities still in the YSO
regime are observed downstream. This situation corroborates the star
formation sequence described by \cite{lon13} and \cite{hen16}.

At isotropic luminosities $\lesssim10^{-6}$\,L$_\odot$, however, the
number of sources increases drastically and in total we observe of
order 600 masers. Their distribution does not follow streamers anymore
but appear more widely distributed (see also \cite[Rickert
2017]{ric17}). Frequently, the masers show double-peaked spectral
profiles, which suggests that the majority of them are associated with
evolved stars across the entire disk of the Milky Way.

\end{document}